\documentstyle[12pt]{article}

\newcommand{\nn}{\nonumber\\}\newcommand{\p}[1]{(\ref{#1})}
\begin{document}
\renewcommand{\thefootnote}{\fnsymbol{footnote}}
\thispagestyle{empty}
\begin{flushright}
JINR E2-2000-65, DFPD 00/TH/19 \\
hep--th/0004019
\end{flushright}
\vspace{.5cm}
\begin{center}
{\large\bf N=4 Superconformal Mechanics and the Potential
Structure\\[.2cm] of AdS Spaces} \vspace{1cm}\\
 E.E. Donets${}^a$\footnote{e-mail: edonets@sunhe.jinr.ru},
A. Pashnev${}^b$\footnote{e-mail: pashnev@thsun1.jinr.ru}, V.O.
Rivelles${}^c$\footnote{e-mail: rivelles@fma.if.usp.br}, D.
Sorokin${}^d$\footnote{ On leave from Institute for Theoretical
Physics, NSC ``Kharkov Institute of Physics and Technology'',
Kharkov 61108, Ukraine. E-mail: dmitri.sorokin@pd.infn.it} and M.
Tsulaia${}^b$\footnote
{e-mail: tsulaia@thsun1.jinr.ru}\vspace{.1cm}\\

\vspace{1.5cm}
${}^a${\it Laboratory of High Energies, JINR}\\
{\it Dubna, 141980, Russia}\\
~\\
${}^b${\it Bogoliubov Laboratory of Theoretical Physics, JINR} \\
{\it Dubna, 141980, Russia}\\
~\\
${}^c${\it Instituto de Fisica,
Universidade de S\~ao Paulo\\
C.Postal 66318, 05315-970 S.Paulo, SP, Brazil}\\
~\\
${}^d${\it Universit\`a Degli Studi Di Padova,
Dipartimento Di Fisica ``Galileo Galilei''\\
ed INFN, Sezione Di Padova
Via F. Marzolo, 8, 35131 Padova, Italia}\vspace{.5cm}\\
{\bf Abstract}
\end{center}
The dynamics of an $N=4$ spinning particle in a curved background
is described using the $N=4$ superfield formalism. The
$SU(2)_{local}\times SU(2)_{global}$ $N=4$ superconformal symmetry
of the particle action requires the background to be a real
``K\"ahler--like" manifold whose metric is generated by a
sigma--model superpotential. The anti--de--Sitter spaces are shown
to belong to this class of manifolds.

\vfill \setcounter{page}0
\renewcommand{\thefootnote}{\arabic{footnote}}
\setcounter{footnote}0
\newpage
\section{Introduction}
Supersymmetric quantum mechanics which underlies the dynamics of
non--relativistic and relativistic spinning particles and
superparticles is one of the simplest examples of supersymmetric
sigma--models and it has attracted a great deal of attention as a
laboratory for studying problems appearing in more complicated
supersymmetric field and string theories. For instance,
one--dimensional \cite{IKP} and multidimensional
\cite{PS,DP} $N=4$ supersymmetric quantum mechanics
(SUSY QM) can be associated with $N=1$, $D=4$ supersymmetric field
theories (including supergravity) subject to an appropriate
dimensional reduction down to $D=1$.

A recent revival of interest in superconformal mechanics \cite{AP,krivonos}
has been caused, in particular, by an observation made in the
context of the AdS/CFT correspondence conjecture that the dynamics
of a superparticle near the AdS horizon of an extreme
Reissner--Nordstr\"om black hole of a large mass is described by a
superconformal mechanics \cite{CDKKTV}. Applications of
supersymmetric mechanics to the theory of black holes and to other
problems have been reviewed in \cite{bri,P}. In \cite{P}
conditions on geometry of curved backgrounds, in which $N=1,2$ and
4 superconformal invariant models of {\it non-relativistic}
spinning particles can exist, have been studied in the $N=1$
superfield formalism.

Note that, as in the case of superstrings, the superconformal
group of the particle superworldline is an infinite dimensional
subgroup of the group of its superdiffeomorphisms. It becomes
manifest in the worldline superfield formulations of {\it
relativistic} spinning particles
\cite{Howe} and superparticles
\cite{STV}, which can thus be regarded as examples of  quantum
mechanics with manifest superconformal symmetry.

The superconformal invariance and, in more general case,
worldvolume superdiffeomorphisms impose restrictions on the
geometry of the background also in models of relativistic
particles and branes. For instance, in the case of superbranes it
requires that a target--superspace background obeys superfield
supergravity constraints (see \cite{Sorokin} for a recent review).

In the case of spinning particles this problem is
connected with the problem of selfconsistent field theoretical description
of interacting particles with spin higher than $2$. It is well known that
the theory of interacting higher spin fields should be formulated in an
anti--de--Sitter background (see \cite{Vasiliev} for a review).

In \cite{HPPT} it was shown that difficulties with constructing a
model of a spinning particle moving in a gravitational background
arise already for spins $3/2$ and  $2$. These difficulties have been overcome
in \cite{kuzenko}, where an action for spinning particles with spin
higher than one were constructed  in backgrounds of
constant curvature (such as the AdS spaces)\footnote{We thank
Sergey Kuzenko for bringing these papers to our attention.}.

The spin 2 particle model of
\cite{GT,HPPT,kuzenko} is based on a so called ``large'' $N=4$
superconformal algebra containing $SO(4)$ as the subalgebra of
local internal symmetries.
It is well known that there exists another (so called ``small'')  $N=4$
superconformal algebra with $SU(2)$ as the subalgebra of local
internal symmetries.
It is therefore tempting to study whether in a superfield formulation of a
spin 2 particle dynamics, which is manifestly invariant under ``small'' $N=4$
superconformal symmetry, conditions imposed on a curved background can be
less restrictive than in the case of \cite{GT,HPPT,kuzenko}.

In this paper we present results of this study.  We consider
relativistic spinning particle mechanics invariant under local
$N=4$ supersymmetry with $SU(2)_{local}\times SU(2)_{global}$
internal symmetries, which is associated with  the ``small'' $N=4$
superconformal algebra. In a flat background this model has been
constructed and studied in \cite{PS}. It was shown that (in four
dimensions) its first quantized spectrum consists of three scalar
and one spin $2$ states corresponding to the linearized limit of a
conformal gravity model. The superfield action for this $N=4$
spinning particle is a localized (or superconformal) version of
the action for $N=4$ supersymmetric quantum mechanics \cite{IKP,
DP} with a quadratic superpotential. This correspondence prompts
us how to generalize the free $N=4$ superconformal spinning
particle action to the description of a particle propagating in a
gravitational background. For this one should consider
supersymmetric quantum mechanics with an arbitrary superpotential
\cite{DP} and make it invariant under the $N=4$ superconformal
transformations \cite{PS}.

In \cite{DP} it has been shown that the $N=4$ superfield formulation
of multidimensional $N=4$ SUSY QM leads to a
supersymmetric nonlinear sigma--model with a target--space
metric being a second derivative of a single real--valued function
(superpotential) $A(x)$
\begin{equation}\label{1}
g_{MN}(x)={{\partial^2A(x)}\over{\partial x^M\partial x^N}} ,
\end{equation}
i.e. for an arbitrary dimension $D$ and signature of the
sigma--model manifold,  parametrized by {\it real} scalar fields
$x^M$ $(M=0,1,...,D-1)$, its metric should have a
``K\"ahler--like" structure. The metric of a similar type appeared
also as a metric of black hole moduli spaces  considered
recently in \cite{gut}.

As has been announced in \cite{PS}, the
$N=4$ superconformal generalization of the model of \cite{DP} in a manifold of
Minkowski signature describes a relativistic spinning particle
propagating in a gravitational background with the metric
(\ref{1}).

It has been known for a long time that supersymmetry requires
sigma--model manifolds of chiral superfields to be K\"ahler,
hyper--K\"ahler \cite{zum}, special K\"ahler \cite{WP}, \cite{Fr}
 or special
Lagrangian manifolds \cite{HG}. The geometrical structure of these
manifolds has been under intensive investigation because of its
relation to the compactification of string theory on Calabi--Yau
manifolds and to duality symmetries of corresponding supergravity
models (see \cite{fre} for a review).

The essential difference of the metric (\ref{1}) from a K\"ahler
metric
\begin{equation}\label{2}
 g_{MN}(z,\bar z)={{\partial^2K(z,\bar z)}\over{\partial
z^M\partial \bar z^N}}
\end{equation}
is that the latter is a Hermitian metric on a complex manifold,
while the former is a {\it real manifold} metric. The reason why a
real sigma--model manifold appears in the case of $N=4$ SUSY QM
under consideration is that we construct the supersymmetric sigma
model with the use of constrained {\it real} superfields and not
with chiral ones as one usually do.

Some K\"ahler manifolds mentioned above also admit real-valued
representation for the metric (\ref{1}). For example, this is so
for a metric of the special K\"ahler manifolds in a flat Darboux
coordinate system \cite{Fr,HG}. However, the class of the
manifolds with the metric (\ref{1}) is more general and includes
manifolds which do not have complex structure.

In particular, we have found that in a certain coordinate system
the metric on an anti--de--Sitter space of an arbitrary dimension
$D$ ($AdS_D$) can be represented in the form (\ref{1}). Other
examples are hyperbolic manifolds of negative curvature on which
M--theory and string theories can compactify \cite{mat,rus}. To
the best of our knowledge this observation is a novel one. This
result can presumably be useful for better understanding the
structure of string and supergravity theories in AdS
superbackgrounds and AdS/CFT correspondence.

The paper is organized as follows. In Section 2 we review the
$N=4$ superconformal particle model of ref. \cite{PS}. In Section
3 we generalize it to describe a spinning particle in a curved
background with the metric (\ref{1}). Properties of the $AdS_D$
space which follow from the potential structure of its metric are
considered in Section 4. In Conclusion we discuss open problems
and outlook.

\section{The free $N=4$ superconformal particle model}
\setcounter{equation}0
We begin with a brief description of free spinning particle
mechanics with $SU(2)_{local}\times SU(2)_{global}$ $N=4$
superconformal symmetry  on the superworldline \cite{PS}. To avoid
confusion we should note that in one-- and two--dimensional
spaces the (super)conformal symmetry is infinite dimensional (i.e.
the parameter of (super)conformal transformations is a
holomorphic function of (super)worldsheet coordinates). The $N=4$
superconformal superalgebra with the local internal  $SU(2)$
automorphisms contains {\em four} supercharges  and is a
subalgebra of a more general
$N=4$ superconformal algebra with an internal local $SO(4)$
which contains {\em eight} supercharges \cite{ademo}. Thus, in our
case $N=4$ counts {\it all} supercharges,  while usually (in
particular in higher dimensions) it corresponds only to
super--Poincare charges and does not include the number of
special superconformal generators.

To construct the superfield action in the worldline superspace
$(\tau,\theta^a, \bar\theta_a)$ (with $\tau$ being a time
parameter, and $\theta^a$ and $\bar\theta_a=(\theta^a)^*,
~(a=1,2)$ being two complex (or four real) Grassmann--odd
coordinates) one introduces $D$ real ``matter'' superfields
$\Phi^M(\tau,\theta^a, \bar\theta_a)$ ($M=0,1,\dots,D-1)$ and a
worldline supereinbein $E(\tau,\theta^a, \bar\theta_a)$ which have
the following properties with respect to the $SU(2)$ $N=4$
superconformal transformations of the worldline superspace
\footnote{Our conventions for spinors are as follows:
${\theta}_{a} ={\theta}^{b}{\varepsilon}_{ba},\; {\theta}^{a}=
{\varepsilon}^{ab}{\theta}_{b},\; {\bar \theta}_{a}={\bar
\theta}^{b}{\varepsilon}_{ba},\; {\bar \theta}^{a}=
{\varepsilon}^{ab}{\bar \theta}_{b},\; {\bar \theta}_a =
(\theta^a)^\ast,\; {\bar \theta}^a = -(\theta_a)^\ast,\; (\theta
\theta)\equiv \theta^a \theta_a = -2 \theta^1 \theta^2,\;
 (\bar \theta \bar \theta)\equiv \bar \theta_a \bar \theta^a = (\theta
\theta)^\ast,\;  (\bar \theta \theta) \equiv \bar \theta_a \theta^a,\;
 \varepsilon^{12}=- \varepsilon^{21}= 1,\;
\varepsilon_{12} = 1$.}
\begin {eqnarray}
\delta \tau&=&\Lambda-\frac{1}{2}\theta^aD_a\Lambda - \frac{1}{2}
\overline{\theta}_a\overline{D}^a\Lambda,\nn
\delta\theta^a&=&i\overline{D}^a\Lambda,\;\;\;
\delta\overline{\theta}_a=iD_a\Lambda, \label{321}\\
 \delta\Phi^M&=&-\Lambda\dot\Phi^M+ \dot\Lambda\Phi^M- i(D_a\Lambda)
(\overline{D}^a\Phi^M)-i(\overline{D}^a\Lambda)( D_a\Phi^M),\\
\delta E&=&-\Lambda\dot{E}-\dot\Lambda E-i(D_a\Lambda)(
\overline{D}^a E)-i(\overline{D}^a\Lambda) (D_a E)\label{TRANS},
\end{eqnarray}
where dot denotes the time derivative $\frac{d}{d\tau}$. The
transformation law \p{TRANS} for the superfields $\Phi^M$ shows
that these superfields are vector superfields in the one
dimensional $N=4$ superspace, while the superfields $E\Phi^M$ are
scalars.

The superfields $\Phi^M$ and $E$ obey the quadratic constraints
\begin{eqnarray} \nonumber
[ D_a , \overline {D}^a]\Phi^M &=&0 , \\ \nonumber D^a D_a \Phi^M &=&0 ,
\\ \label{C1} \overline{D}_a \overline{D}^a \Phi^M &=&0 ,
\end{eqnarray}
and
\begin{eqnarray} \nonumber
[ D_a , \overline{D}^a]\frac{1}{E} &=&0 , \\ \nonumber
D^a D_a \frac{1}{E} &=&0 , \\ \label {C2}
\overline {D}_a \overline {D}^a \frac{1}{E} &=&0 ,
\end{eqnarray}
where
\begin{equation}
D_a = \frac{\partial}{\partial \theta^a}- \frac{i}{2} \overline{\theta}_a
\frac{\partial}{\partial \tau} ,
\quad
\overline{D}^a = \frac{\partial}{\partial \overline{\theta}_a}
- \frac{i}{2} \theta^a
\frac{\partial}{\partial \tau} ,
\end{equation}
are the supercovariant derivatives, and the infinitesimal
superfield
\begin{eqnarray}
\Lambda(\tau,\theta,\overline{\theta})&=&a(\tau)
+\theta^a \overline{\alpha}_a(\tau)
-\overline{\theta}_a\alpha^a(\tau)
+{\theta^a}(\sigma_i)^{~b}_a  \overline{ \theta}_b b_i(\tau)
\nn
\label{322} &&-\frac{i}{2}(\theta^a \dot{\overline{\alpha}}_a(\tau) +
\overline{\theta}_a \dot\alpha^a(\tau)) \overline{\theta}\theta+
\frac{1}{8}(\overline{\theta}\theta)^2\ddot{a}(\tau)
\end{eqnarray}
contains the parameters of local reparametrizations $a(\tau)$,
local supertranslations $\alpha^a(\tau)$, $\overline \alpha_a(\tau)$
 and local $SU(2)$
rotations $b_i(\tau)$ of the worldline superspace. It is
constrained by the same relations \p{C1} as $1/E$ and $\Phi^M$
(($\sigma_i)^{~b}_a$ are the Pauli matrices, $i=1,2,3$).

The constraints \p{C1}--\p{C2} can be explicitly solved \cite{PS, MU},
the solution being described by the superfields
\begin{eqnarray}
\Phi^M(\tau,\eta,\overline{\eta})&=&{1\over {e(\tau)}} x^M(\tau)
+\theta^a \overline{\psi}'^M_a(\tau)
-\overline{\theta}_a \psi'^{Ma}(\tau)+
{\theta}^a(\sigma_i)^{~b}_a {\overline \theta}_b T'^M_i(\tau)
\nn \label{327}
&&-\frac{i}{2}(\theta^a \dot{\overline{\psi}}_a^{\prime M}(\tau)
+\overline{\theta}_a\dot\psi'^{Ma}(\tau)) \overline{\theta}\theta
+\frac{1}{8}(\overline{\theta}\theta)^2{d^2\over{d\tau^2}}
\left({1\over {e(\tau)}}{x}^M\right),
\end{eqnarray}
and
\begin{eqnarray}
\frac{1}{E}(\tau,\theta,\overline{\theta})&=&\frac{1}{e(\tau)}
+\theta^a \overline{\lambda}_a'(\tau)
-\overline{\theta}_a\lambda'^a(\tau)
+\theta^a (\sigma_i)^{~b}_a{\overline \theta}_b
t'_i(\tau) \nn \label{326}
&&-\frac{i}{2}
(\theta^a \dot{\overline{{\lambda}'}}_a(\tau)
+\overline {\theta}_a \dot\lambda'^a(\tau))
\overline{\theta}\theta+
\frac{1}{8}(\overline{\theta}\theta)^2\frac{d^2}{d\tau^2}\frac{1}{e(\tau)}.
\end{eqnarray}
The leading components $x^M(\tau)$ of the superfields $\Phi^M$ are
associated with coordinates of the particle trajectory in a
$D$--dimensional flat target space--time, the Grassmann--odd
vectors $\psi'^{Ma}(\tau)$ and $ \overline{\psi'}^M_a(\tau)$
correspond to particle spin degrees of freedom and $T'^M_i(\tau)$
are auxiliary fields. The superfield $1/E$ describes an $N=4$
worldline supergravity multiplet consisting of the einbein
(``graviton") $e(\tau)$, two complex ``gravitini" $\lambda'^a(\tau)$
and $\overline\lambda'_a(\tau)$, and the $SU(2)$ gauge field
$t'_i(\tau)$. Upon an appropriate field redefinition (see eqs.
(\ref{def1}) of the next section) we shall pass from ``primed" to
``unprimed" component fields.

The $N=4$ superfield action for a relativistic spinning particle
in a flat target space has the following form \cite{PS}
\begin{equation} \label{LAGR_PS}
S=-8 \int d\tau d^2\theta d^2 \overline{\theta} E\Phi^M\Phi^N
\eta_{MN}
\end{equation}
where  $\eta_{MN}=diag~(-,+,\dots,+)$ is the Minkowski metric. The
components of $E$ play the role of Lagrange multipliers. Their
presence implies that the dynamics of the particle is subject to
relativistic constraints, in particular, the particle is
massless $(p_Mp^M=0)$ . The Dirac quantization of the model
(\ref{LAGR_PS}) shows that its quantum spectrum consists of one
spin $2$ and three spin $0$ particle states and it can be regarded
as a linearized spectrum of a conformal gravity \cite{PS}.

\section{The spinning particle in a curved background}
\setcounter{equation}0
 Let us now generalize the model of the previous section to describe
a spinning particle propagating in the gravitational background.
To this end we replace \p{LAGR_PS} with the most general action
functional which respects the $N=4$ superconformal symmetry
\begin{equation} \label{LAGR}
S=-8 \int d\tau d^2\theta d^2 \overline{\theta} E^{-1}A(E\Phi^M),
\end{equation}
where $A(E\Phi^M)$ is an arbitrary function (called the
superpotential) of $E\Phi^M$. Recall that $E\Phi^M$ transform as
scalar superfields with respect to (\ref{TRANS}), while $\Phi^M$
and $1\over E$ are vectors. Note also that $E^{-1}A(E\Phi^M)$ can
be regarded as a rank one homogeneous function in a $D+1$
dimensional space with $x^D=E^{-1}$. A consequence of such a
structure of the superfield action \p{LAGR} is the fact that only
$D$ of the bosonic coordinates in the $D+1$ dimensional space
describe dynamical degrees of freedom. The einbein $e(\tau)$ and
its superpartners are auxiliary fields as in the free particle
case \p{LAGR_PS}.

Integrating \p{LAGR} over the Grassmann coordinates $\theta^a$,
$\overline{\theta}_a$ and making the following redefinition of the
component fields
$$ \lambda^a=e^{\frac{3}{2}}\lambda'^a, \quad \overline {\lambda}_a=
(\lambda^a)^\ast, \quad
t_i=2e({t'}_i+ e{\lambda'^b(\sigma_i)^{~a}_b\overline{\lambda}}'_a),  \quad
\psi^{Ma}=\sqrt{e}(\psi'^{Ma}- x^M \lambda'^a),
$$
\begin{equation}\label{def1}
\overline {\psi}^M_a =(\psi^{Ma})^\ast,  \quad
T_i^M=2\sqrt{e} (T'^M_i-  x^M t'_i
+\frac{\sqrt{e}}{2}\lambda'^b  (\sigma_i)^{~a}_b \overline{\psi}^M_a
+\frac{\sqrt{e}}{2} {\psi^{Mb} (\sigma_i)^{~a}_b\overline{\lambda}}'_a),
\end{equation}
one obtains the component action
\begin{equation} \label{COMP}
S = \int d\tau(K - V) ,
\end{equation}
where
\begin{eqnarray}\label{K} \nonumber
K&=&\frac{1}{2e}g_{MN}
(\dot x^M -i\overline {\lambda}_a \psi^{Ma}
+i \overline {\psi}^M_a \lambda^a)
(\dot x^N -i\overline {\lambda}_b \psi^{Nb}
+i \overline {\psi}^N_b \lambda^b) \\
&& + i g_{MN} (\overline {\psi}_a^M \dot \psi^{aN} +
\psi^{aM} \dot{\overline {\psi}_a^N})
\end{eqnarray}
is the kinetic term and
\begin{eqnarray}\label{V}
V&=& - \frac{1}{2}g_{MN} T^M_i T^N_i + 2\sqrt{e} \Gamma_{LMN}
\psi^{Mb}(\sigma_i)^{~a}_b \overline {\psi}_a^{L} T^N_i - t_i g_{MN}
\psi^{Nb} (\sigma_i)^{~a}_b\overline \psi^{M}_a \\ \nonumber
&&+2\Gamma_{LMN} (\lambda^a
\overline {\psi}^L_a \overline {\psi}^M_b \psi^{bN} +
\overline {\lambda}_b \psi^{Lb}
\psi ^{Ma} \overline {\psi}_a^{N}) +e (\partial_L\Gamma_{MNP})
 (\overline {\psi}^L_a \overline {\psi}^{Ma})(\psi^{Nb} \psi^P_b)
\end{eqnarray}
describes fermionic interactions.
 In eqs. \p{K} and \p{V}
 \begin{equation}\label{metric}
g_{MN}(x)={{\partial^2}\over{\partial x^M\partial x^N}}
A(x)\equiv \partial_{MN}^2 A(x), \quad
A(x^M)=A(E\Phi^M)|_{\theta,\overline \theta=0}
 \end{equation}
is the metric of a sigma--model $D$--dimensional manifold
parametrized by the worldline scalar fields $x^M(\tau)$ and
 \begin{equation}\label{chris}
\Gamma_{LMN}(x) = {1\over 2}\partial^3_{LMN} A(x)
 \end{equation}
is the (totally symmetric) Christoffel connection associated with
$g_{MN}$ (i.e. ${\cal
D}_Lg_{MN}=\partial_Lg_{MN}-\Gamma_{LM}^{~~P}g_{PN}-\Gamma_{LN}^{~~P}g_{PM}=0$).

The Riemann curvature of this manifold has the form
\begin{equation}\label{curv}
R_{LM,NP}=\Gamma_{LP}^{~~~Q}\Gamma_{QMN}-\Gamma_{LN}^{~~~Q}\Gamma_{QMP}.
\end{equation}

Upon solving for the equations of motion of the auxiliary fields
$T^M_i$, substituting the solution back into eqs. \p{COMP}--\p{V}
and performing Legendre transformations one arrives at the first
order form of the spinning particle action
\begin{equation}\label{fo}
S=\int d\tau\left[p_M\dot
x^M+i(\psi_M^a \dot{\overline {\psi}_a^M}
+ \overline {\psi}_{Ma} \dot\psi^{Ma})-H\right],
\end{equation}
where $p_M$ is the momentum canonically conjugate to $x^M$, and
the Hamiltonian $H$ of the system has the following structure
\begin{equation}
H = e(\tau) H_{0} + i\lambda^a(\tau) \overline {Q}_{a} +
i\overline {\lambda}_a(\tau) Q^a - t_i(\tau) L_{i},
\end{equation}
with
\begin{eqnarray}\label{h0} \nonumber
H_{0}&=& \frac{1}{2}g^{MN}p_M p_N +R_{LN,PM}
(\overline {\psi}^L_a \overline {\psi}^{Ma})(\psi^{Nb} \psi^{P}_b)
 +R_{MP,NL}(\overline {\psi}^L_{a} \psi^{Ma})
(\overline {\psi}^{N}_b \psi^{Pb})\\
 &&+{\cal D}_LG_{MNP}(\overline {\psi}^L_a \overline {\psi}^{Ma})
(\psi^{Nb} \psi^P_b),
\end{eqnarray}
\begin{equation} \label{CHARGE1}
\overline {Q}_{a} = \overline {\psi}^M_a p_M + i\Gamma_{LMN}
\overline {\psi}^L_c \overline {\psi}^{Mc} \psi_a^N,
\end{equation}
\begin{equation} \label{CHARGE2}
Q^b=\psi^{Ib} p_I + i\Gamma_{LMN}\overline {\psi}^{Lb} \psi^{Mc} \psi_c^N
\end{equation}
and
\begin{equation}\label{su2}
L_{i} = g_{MN} \psi^{Nb} (\sigma_i)^{~a}_b  \overline {\psi}^{M}_a
\end{equation}
being associated with constraints on the dynamics of the
relativistic particle caused by the worldline
superreparametrization invariance of the model. The constraints
are of the first class since they form a closed $N=4$
supersymmetry algebra $$ \{ \overline {Q}_a , Q^b \} = -i \delta^b_a
H_{0} ,  \quad [ L_i , L_j ] = \epsilon_{ijk} L_k , $$
\begin{equation}\label{algebra}
[L_i , \overline {Q}_a ] =  \frac{i}{2} {(\sigma_i)}^{~c}_a
\overline {Q}_c ,
\quad [ L_i ,  Q^a ] = - \frac{i}{2} {(\sigma_i)}^{~a}_c  Q^c
\end{equation}
 with respect to the following graded Dirac brackets \cite{DP}
 (which are obtained upon solving for the second class constraints
 on the canonical fermionic momenta $\pi_{Ma}= - i\overline {\psi}_{Ma}$
and  $\overline {\pi}^a_M=- i{\psi}^a_M$
 derived from eq. \p{fo})
$$ [x^M, p_N ] =  \delta^M_N, \quad \{ \psi^{aM}, \overline {\psi}^N_b \}
=- \frac{i}{2}\delta^a_b  g^{MN} , \quad [p_M,p_N]=2iR_{MN,PL}
\overline{\psi}_a^P\psi^{aL}, $$
\begin{equation}
[ p_M,\psi^{a}_N]=\Gamma_{MNP} \psi^{aP}, \quad
[p_M,\overline {\psi}^{a}_N]=\Gamma_{MNP} \overline{\psi}^{aP},
\end{equation}
We observe that $p_M$ have properties of covariant momenta when
acting on fermionic variables $\psi^{Ma}$ and $\overline {\psi}^M_a$.

The superalgebra \p{algebra} of the constraints \p{h0}--\p{su2}
generates the $SU(2)_{local}\times SU(2)_{global}$ $N=4$
superconformal transformations \p{TRANS} of the components of the
superfields $\Phi^M$.

We have thus shown that the $N=4$ worldline superfield action
\p{LAGR}, which reduces to \p{fo}--\p{su2} upon integrating over
Grassmann--odd coordinates and eliminating auxiliary fields,
describes the dynamics of an $N=4$ superconformal spinning
particle in a curved background whose geometry is characterized
by eqs. \p{chris}--\p{curv}.

We should note that the last terms in \p{h0}--\p{CHARGE2},
containing the Christoffel connection, are non-covariant with
respect to general coordinate transformations of the background.
The reason is that background diffeomorphisms acting on the
superfields $\Phi^M$, in general, are incompatible with the
constraints \p{C1}--\p{C2}. This, in particular, means that if a
background metric \p{metric} admits isometries, not all of them
will be symmetries of the actions \p{LAGR} and \p{fo}. It is an
interesting open problem to study whether the model under
consideration can be modified in such a way that only
target--space covariant terms remain in the action.

\section{The potential structure of the anti--de--Sitter metric}
\setcounter{equation}0

It is curiously enough that the anti--de--Sitter spaces belong to
the class of the manifolds whose metric in a certain coordinate
system acquires the form \p{1}. To show this consider first a
coordinate system
\begin{equation}\label{cf}
X^M=(X^\mu, \rho), \quad \mu = 0,..., D-2
\end{equation}
in which the metric of a $D$--dimensional AdS space has a
conformally flat form (for simplicity we put the AdS radius to
one)
\begin{equation} \label{mc}
ds^2 = \frac{1}{\rho^2}\left(\eta_{\mu \nu}dX{^\mu} dX^{\nu} +
d\rho^2\right),
\end{equation}
 where
$\eta_{\mu \nu} = (-1,1, ..., 1)$.

Now perform a coordinate transformation to the new set of variables
\begin{equation}\label{new}
x^{M}=(x^\mu, r)
\end{equation}
such that
\begin{equation} \label{AdS1}
X^{\mu} = \frac{x^\mu}{r}, \quad \rho = \frac{1}{\sqrt r}.
\end{equation}
The passage from $\rho$ to $r$ has proved to be convenient for the analysis
of the properties of the potential $A(x)$ considered below.

In the coordinate system \p{new} the AdS metric $g_{MN}$ takes the
form
\begin{equation}\label{newm}
g_{\mu\nu} = \frac{\eta_{\mu \nu}}{r}, \quad g_{\mu r} = -
\frac{\eta_{\mu \nu} x^\nu}{r^2},
 \quad g_{rr}=  \frac{x^\mu x^\nu\eta_{\mu\nu}}{r^3}
+\frac{1}{4r^2},
\end{equation}
where the index $r$ of the metric tensor components corresponds to the
coordinate $r$.

One can easily check that the metric \p{newm} is a second
derivative of the following function
\begin{equation}\label{SU1}
A(x) = \frac{x^\mu x^\nu \eta_{\mu \nu}}{2r}- \frac{1}{4}\ln r.
\end{equation}
Thus we have shown that $AdS_D$ is one of the manifolds of the
type \p{1}, where the $N=4$ superconformal spinning particle can
live.

By passing  note that if in the action \p{LAGR} we take
$A(E\Phi^M)$ in the form \p{SU1} and put $\Phi^\mu=0$ and
$\Phi^r=1$ we shall arrive at the action
 $$S=2\int
d\tau d^2\theta d^2\bar\theta~{{\ln E}\over E}
 $$
 which describes a one--dimensional $N=4$ superconformal mechanics considered in
\cite{krivonos}.

The potential \p{SU1} generating the metric on $AdS_D$ is not
unique. Another form of the potential arises when one performs the
following change of variables \p{cf}
\begin{equation} \label{AdS3}
X^{\mu} =\left( \frac{x^\mu}{r}\right)^{m_\mu} \quad \rho =
\frac{1}{\sqrt r},
\end{equation}
where $m_\mu\not = 0,{1\over 2}$ is a set of real numbers, namely
\begin{equation}\label{SU3}
A = -\frac{m^2_0}{2m_0(2m_0 -1)} \frac{(x^0)^{2m_0}}{r^{2m_0-1}} +
\sum_{i=1}^{D-2}\frac{m^2_i}{2m_i(2m_i -1)}
\frac{(x^i)^{2m_i}}{r^{2m_i-1}}
- \frac{1}{4}\ln r.
\end{equation}

More generally, we could  make, for instance, a ``logarithmic
transformation"
\begin{equation}\label{AdS2}
X^{\mu} = \ln (\frac{x^\mu}{r}) \quad \rho = \frac{1}{\sqrt r},
\end{equation}
for which the corresponding potential has the form
\begin{equation}\label{SU2}
A=r \ln \frac{x^0}{r} -
 \sum_{i=1}^{D-2} r \ln \frac{x^i}{r}
- \frac{1}{4} \ln r.
\end{equation}

The coordinate transformation \p{AdS1} is singled out by the
requirement that it is a single--valued and that a Lorentz
subgroup $SO(1,D-2)$ of the $AdS_D$ isometry group $SO(2,D-1)$
acts linearly on both the ``old" coordinates $X^\mu$ of \p{cf} and
the ``new" coordinates $x^{\mu}$ of \p{new}, \p{AdS1}.

 So we shall further discuss some amusing properties of $AdS_D$
associated with its potential structure \p{SU1} in the coordinate
system \p{new}.

The group $SO(2,D-1)$ of the isometry transformations of $AdS$
coordinates, which leave the form of the $AdS$ metric invariant,
is known to act as a conformal group on a $(D-1)$-dimensional
boundary of $AdS_D$. In the coordinate system \p{cf} the boundary
(which is a $(D-1)$--dimensional Minkowski space) is associated
with the coordinates $X^\mu$. Under infinitesimal $SO(2,D-1)$
transformations $X^\mu$ and $\rho$ vary as follows
\begin{eqnarray}\label{cf1} \nonumber
\delta X^\mu &=& a^\mu+a^{\mu\nu} X^\lambda \eta_{\nu \lambda}
+a_D X^\mu+a^\mu_K X^\nu
X^\lambda \eta_{\nu \lambda}
-2(a^\nu_KX^\lambda \eta_{\nu \lambda})X^\mu +{a^\mu_K\rho^2}, \\
\delta \rho &=&-(2a^\mu_K X^\nu \eta_{\mu \nu}-a_D)\rho,
\end{eqnarray}
where the $SO(2,D-1)$ parameters
 $
a^{\mu},~a^{\mu\nu}, ~a_D,~ a^{\mu}_K
 $
are, respectively, the parameters of $D-1$ translations,
$SO(1,D-2)$ rotations, dilatation and conformal boosts, acting as
conformal transformations in a $(D-1)$--dimensional slice of
$AdS_D$ parametrized by $X^\mu$.

  From \p{AdS1} and \p{cf1} one gets the infinitesimal $SO(2,D-1)$
transformations of $x^\mu$ and $r$ \p{new}
\begin{eqnarray}\label{new1} \nonumber
\delta x^\mu&=&a^\mu r+a^{\mu\nu}x^\lambda \eta_{\nu \lambda}
-a_D x^\mu+a^\mu_K {{x^\nu
x^\lambda \eta_{\nu \lambda}}\over r}+
2(a^\nu_Kx^\lambda \eta_{\nu \lambda}){x^\mu\over r} +a^\mu_K,\\
\delta r&=&4a^\mu_Kx^\nu \eta_{\mu \nu}-2a_Dr.
\end{eqnarray}
Under \p{new1} the potential \p{SU1} varies as follows
 $$
 A(x')=A(x)+\delta A(x),
 $$
\begin{equation}\label{dA}
\delta A(x)=\delta x^M\partial_MA(x)=a^\mu x^\nu \eta_{\mu \nu}+{1\over 2}
a_D+a^\mu_Kx^\nu\eta_{\mu \nu}
 {{x^\lambda x^\sigma \eta_{\lambda \sigma}}\over r^2}.
\end{equation}

One can check that the form of the metric \p{AdS1} remains
invariant under the action of the $SO(2,D-1)$ transformations
\p{new1}, so that they are indeed the isometries of this $AdS$
metric. However, the superfield action \p{LAGR} is invariant only
under the subgroup of $SO(2,D-1)$ generated by $D-1$ translations
$a^\mu$, $SO(1,D-2)$ Lorents rotations $a^{\mu\nu}$ and
dilatations $a_D$ which transform the superfields $E\Phi^M$ in the
same way as $x^M$ in \p{new1}. As can be seen from the form of the
variation of $x^M$ (and respectively of $E\Phi^M$) with respect to
conformal boosts $a^\mu_K$, the corresponding term does not
satisfy the superfield constraints \p{C1} and \p{C2}, and, hence,
the transformed $\Phi^M$ will not do so as well. This is the
reason of the appearance of noncovariant terms depending on the
Christoffel connection in the component actions  \p{COMP} and
\p{fo}--\p{su2}.

An interesting property of the potential \p{SU1} is that the
contraction of its partial derivatives with the coordinates
\p{new} are constants starting from the second derivative
 $$
 x^Mx^N\partial^2_{MN}A(x)=x^Mx^Ng_{MN}={1\over 4}
$$
 $$
 x^Lx^Mx^N\partial^3_{LMN}A(x)=2x^Lx^Mx^N\Gamma_{LMN}=-{1\over 2}
 $$
 \begin{equation}\label{series}
 x^{M_1}...x^{M_{n+1}}\partial^{n+1}_{M_1\dots M_{n+1}}A(x)
 =(-1)^{n+1}{{n+1}\over 4}, \quad n=1,\dots, \infty .
 \end{equation}

To get the relation \p{series} one should note that under the
following rescaling of the coordinates \p{new} $x^M\rightarrow
(1+\epsilon)x^M$ (where $\epsilon$ is a numerical
parameter)\footnote{One should not confuse this rescaling with
dilatation isometry  \p{new1} which acts as follows
$x^\mu\rightarrow (1+\epsilon)x^\mu$ and $r\rightarrow
(1+\epsilon)^2r$.} the potential \p{SU1} takes the form
\begin{eqnarray}
A_\epsilon &=& (1+\epsilon)\frac{x^\mu x^\nu \eta_{\mu \nu}}{2r}
- \frac{1}{4}\ln
r- \frac{1}{4}\ln{(1+\epsilon)} \nonumber \\ \label{Ae}
 &=& (1+\epsilon)\frac{x^\mu x^\nu \eta_{\mu \nu}}{2r}
- \frac{1}{4}\ln r-{1\over
4}\epsilon +{1\over
4}\sum_{n=1}^{\infty}(-1)^{n+1}{{n+1}\over{(n+1)!}}\epsilon^{n+1},
\end{eqnarray}
where on the right hand side of \p{Ae} we have expanded
$\ln{(1+\epsilon)}$ in series of $\epsilon$.

On the other hand
\begin{equation}\label{Ae2}
A_\epsilon=A(x+\epsilon x)=A(x)+\epsilon x^M\partial_MA(x)+
\sum_{n=1}^{\infty}{{\epsilon^{n+1}}\over{(n+1)!}}
x^{M_1}...x^{M_{n+1}}\partial^{n+1}_{M_1\dots M_{n+1}}A(x).
\end{equation} Comparing \p{Ae} with \p{Ae2} we get \p{series}.

A local basis in a tangent space of the $AdS_D$ manifold can be
described by the following vielbein one--form
$e^A=dx^Me_M^{~A}(x)$ ($A=0,1\dots,D-1$)
\begin{equation}\label{viel}
e^\alpha=dx^\mu~\delta^\alpha_\mu r^{-{1\over 2}}, \quad
e^r=-dx^\mu ~x_\mu r^{-{3\over 2}}+dr{1\over 2r}
\end{equation}
determined such that $g_{MN}=e^{~A}_Me^{~B}_N\eta_{AB}$ and
$\eta_{AB}=(-,+\dots,+)$.

 Using \p{viel} it is easy to calculate the determinant of the metric \p{AdS1}
\begin{equation}\label{det}
\det g_{MN}=-(\det e_M^{~A})^2=-{1\over{4r^{D+1}}}.
\end{equation}

One more observation concerns the form of the covariant
derivative of the $AdS_D$ Christoffel connection \p{chris}
appeared in \p{h0}. A direct computation results in the following
relation
\begin{equation}\label{dgamma}
{\cal D}_L\Gamma_{MNP}={1\over
8}\partial^4_{LMNP}A(x)-g_{LM}g_{NP}-g_{MN}g_{LP}-g_{NL}g_{MP}.
\end{equation}
We see that noncovariance of \p{dgamma} is in a certain sense
"concentrated" in the fourth partial derivative of $A(x)$.

Note that the results of this section do not depend on the signature of
the metric $\eta_{\mu\nu}$ in \p{mc}. For instance, we could equally well
choose $\eta_{\mu\nu}$ to be Euclidean. Then we would deal with Euclidean
AdS, or hyperbolic spaces considered recently in the context of string
and M--theory compactifications \cite{mat,rus}.

\setcounter{equation}0
\section{Discussion}

To conclude, in this paper we have considered the classical
dynamics of a spinning particle governed by the action invariant
under the $SU(2)_{local} \times SU(2)_{global}$ $N=4$
superconformal transformations of the particle superworldline. We
have shown that the $N=4$ superconformal invariance allows the
particle to propagate in a curved background with a
``K\"ahler--like" metric generated by a real superpotential $A$,
and we have found that the anti--de--Sitter and hyperbolic
spaces belong to this
class of manifolds.

There are several directions of the extension of the results of
this paper. One of them is the quantum description of the $N=4$
superconformal particle model, which can be carried out following
either the lines of \cite{PS,DP} or using path integral
quantization methods. The latter procedure seems to be more
attractive, since it may lead to deeper understanding of the
model, for instance, in the context of the AdS/CFT correspondence
conjecture.

In particular, it is interesting to study both the classical and
quantum dynamics of the $N=4$ superconformal spinning particle
moving in backgrounds which are direct products of $AdS_D$ and
K\"ahler manifolds. Particle motion on the K\"ahler manifolds can
be described by making a multidimensional generalization of the
$N=4$ supersymmetric quantum mechanics considered in \cite{BPCH}.
For this, in addition to $\Phi^M$, one should introduce a number
of chiral superfields $\Psi^n(\tau,\theta,\bar\theta)$
($\overline{D}^a\Psi^n=0$)
\begin{eqnarray} \nonumber
\Psi^n(\tau,\theta,\overline{\theta})&=&z^n(\tau)+\theta^a\chi^n_a(\tau)+
\frac{i}{2} \overline{\theta}\theta\dot{z}^n(\tau)+ \theta\theta
F^n(\tau)   \\
&&- \frac{i}{4}\theta\theta \overline{\theta}_a\dot{\chi}^{na}
(\tau)-
\frac{1}{16}\overline{\theta}\overline{\theta}\theta\theta\ddot{z}^n(\tau),
\end{eqnarray}
and their complex conjugate antichiral superfields ${\overline
\Psi}^n(\tau,\theta,\overline{\theta})$. The superfields $\Psi$
and $\bar\Psi$ transform as scalars under the $N=4$ superconformal
transformations \p{321}.

We can add to the action \p{LAGR} the following $N=4$
superconformal invariant action constructed from $\Psi^n$ and
$\bar\Psi^n$
\begin{equation}\label{action1}
S_K = 2\int d \tau d^2 \theta d^2 \overline{\theta}
\frac{1}{E}K(\Psi,\overline{\Psi}),
\end{equation}
where $K$ is a K\"ahler superpotential.

When the superpotential $A(E\Phi^M)$ is chosen in the form
\p{SU1}, the sum of the actions (\ref{LAGR}) and (\ref{action1})
describes a spinning particle propagating in an $AdS_D \times
K_{2n}$ background, where $K_{2n}$ is a K\"ahler manifold with a
metric \p{2}.
For instance, the case  $n=1$ and  $K_2=\ln(1 +\Psi
\overline{\Psi})$ corresponds to a two-dimensional sphere $S^2$,
which is known to be a K\"ahler manifold.
A detailed analysis of these models will be given elsewhere.

\noindent {\bf Acknowledgments.} We are grateful to Dmitri
Fursaev, Paul Howe, Armen Nersessian, George Papadopoulos, Paolo
Pasti, Volodya Rubtsov, Mario Tonin and Peter West for interest to
this work and helpful discussions. Work of  A.P. and M.T. was
supported in part by the Russian Foundation of Fundamental
Research, under the grant 99-02-18417 and the joint grant RFFR-DFG
99-02-04022, and by a grant of the Committee for Collaboration
between Czech Republic and JINR. Work of V.O.R. was partially
supported by CNPq and a grant by FAPESP. Work of D.S. was
partially supported by the European Commission TMR Programme
ERBFMPX-CT96-0045 to which the author is associated. M.T. is
grateful to the Abdus Salam International Centre for Theoretical
Physics, Trieste, where a part of this work was done.

\end{document}